\documentstyle[twocolumn,prd,aps,epsfig]{revtex}

\begin{document}
\draft
\topmargin=-.5in

\twocolumn[\hsize\textwidth\columnwidth\hsize\csname
@twocolumnfalse\endcsname

\title{Higgs boson decay to
\boldmath$\mu$\boldmath$\bar\mu$\boldmath$\gamma$}

\author{Ali Abbasabadi}
\address{Department of Physical Sciences,
Ferris State University, Michigan 49307, USA}

\author{Wayne W. Repko}
\address{Department of Physics and Astronomy,
Michigan State University, East Lansing, Michigan 48824, USA}

\date{\today}
\maketitle
\begin{abstract}
\hspace*{0.1cm}
 The Higgs boson decay, $H\rightarrow \mu\bar\mu\gamma$, is
studied in the Standard Model at the tree and one-loop levels. It is shown that
for Higgs boson masses above 110 GeV, the contribution to the radiative width
from the one-loop level exceeds the contribution from the tree level, and for
Higgs boson masses above 140 GeV, it even exceeds the contribution from the
tree level decay $H\rightarrow \mu\bar\mu$. We also show that the contributions
to the radiative decay width from the interference terms between the tree and
one-loop diagrams are negligible.
\end{abstract}

\vskip1pc]

\section{Introduction}

 For intermediate mass Higgs ($H$) bosons, the decay
$H\rightarrow \gamma\gamma$ is usually viewed as the discovery mode \cite{hhg},
although decay modes such as $H\rightarrow Z\gamma$ and $H\rightarrow b\bar{b}$
are often considered. Here, motivated by recent studies of muon colliders
\cite{kaplan}, we examine Higgs decays into muons accompanied by a photon. At
muon colliders, these decays will, at the very least, represent a radiative
correction to the measurement of the width $\Gamma(H\to\mu\bar{\mu})$. For
certain values of the Higgs boson mass, $m_H$, this radiative process can be
quite large due to one-loop corrections. As shown in a previous study of the
decays $H\rightarrow f \bar{f} \gamma$ \cite{ab-cdr1}, where $f$ is a light
fermion, the dominant contributions come from the one-loop level when $m_f$ is
negligible. The one-loop decay channel calculation is related to that of the
scattering process $e\bar{e} \rightarrow H\gamma$ \cite{ab-cdr2,ddhr}. For
muons, the dominance of the one-loop calculation must be reexamined.

Due to the relatively large mass of muon ($m_{\mu}\gg m_e$), the
Higgs-boson-muon coupling is sufficiently large to make the tree level
contribution to the decay $H\rightarrow \mu\bar\mu\gamma$ significant. Our
calculations show that the tree level amplitudes are dominated by the muon
helicity non-flip terms, while the dominant contributions to the one-loop
amplitudes come from the muon helicity flip terms. Therefore, the contributions
to the radiative decay width from the interference terms between the tree and
one-loop diagrams turn out to be negligible.

 In the next section, we present general discussion of the kinematics of the
Higgs boson decay width and muon invariant-mass distribution for $H\rightarrow
\mu\bar\mu\gamma$, and the cuts imposed on the $\mu$, $\bar\mu$, and $\gamma$.
Section III gives the tree level amplitudes for the decay $H\rightarrow
\mu\bar\mu\gamma$ and a summary of the results for the one-loop calculation.
Section IV contains the combined contributions from the tree and one-loop
levels to the decay of the Higgs boson. This is followed by a summary which
includes a discussion of possibility of using the decay $H\to\mu\bar\mu\gamma$
as a probe of the Higgs boson coupling to the top quark.

\section{Higgs Boson Decay Widths}

The muon-invariant mass distribution $d\Gamma/dm_{\mu\bar\mu}$ and the width
$\Gamma$ for the decay $H\rightarrow \mu\bar\mu\gamma$ are given by
\begin{equation}\label{dgam}
\frac{d\Gamma}{dm_{\mu\bar\mu}} =
\frac{1}{128\pi^3}\frac{m_{\mu\bar\mu}}{m_H^3}
\int_{(m_{\mu\gamma}^2)_{\rm min}}^{(m_{\mu\gamma}^2)_{\rm max}}
dm_{\mu\gamma}^2\,\sum_{\rm spin}|{\cal M}|^2\,,
\end{equation}
\begin{equation}\label{gam}
\Gamma = \int_{(m_{\mu\bar\mu}^2)_{\rm min}}^{(m_{\mu\bar\mu}^2)_{\rm max}}
dm_{\mu\bar\mu}^2\,\frac{1}{2m_{\mu\bar\mu}}\frac{d\Gamma}{dm_{\mu\bar\mu}}\,,
\end{equation}
with the Lorentz-invariant amplitude ${\cal M}$ specified in the next section.
The invariant masses $m_{\mu\bar\mu}$ and $m_{\mu\gamma}$ are related to the
Mandelstam variables $s$, $t$, and $u$, by $s = m_{\mu\bar\mu}^2 =
(p_{\mu}+p_{\bar\mu})^2$, $t = m_{\mu\gamma}^2 = (p_{\mu}+p_{\gamma})^2$, and
$u = m_{\bar\mu\gamma}^2 = (p_{\bar\mu}+p_{\gamma})^2$. Here, $p_{\mu}$,
$p_{\bar\mu}$, and $p_{\gamma}$ are the 4-momenta for the $\mu$, $\bar\mu$, and
$\gamma$, respectively. The limits on the $dm_{\mu\gamma}^2$ and
$dm_{\mu\bar\mu}^2$ integrations, without imposing any cuts, are
\begin{eqnarray}
(m_{\mu\gamma}^2)_{\rm min} & = &\;m_\mu^2 +
\case{1}{2}(m_H^2 - m_{\mu\bar\mu}^2)\left(1 - \beta\right)\,, \\
(m_{\mu\gamma}^2)_{\rm max} & = &\;m_\mu^2 +
\case{1}{2}(m_H^2 - m_{\mu\bar\mu}^2)\left(1 + \beta\right)\,, \\
(m_{\mu\bar\mu}^2)_{\rm min} & = &\;4m_\mu^2\,,\\
(m_{\mu\bar\mu}^2)_{\rm max} & = &\;m_H^2\,,
\end{eqnarray}
with $\beta = \sqrt{1 - 4m_\mu^2/m_{\mu\bar\mu}^2}$. Because the full phase
space is not experimentally accessible, we impose the following cuts:
$m_{\mu\bar\mu}^2\geq(m_{\mu\bar\mu}^2)_{{\rm cut}}$,
$m_{\mu\gamma}^2\geq(m_{\mu\gamma}^2)_{{\rm cut}}$,
$m_{\bar\mu\gamma}^2\geq(m_{\bar\mu\gamma}^2)_{{\rm cut}}$,
$E_\mu\geq(E_\mu)_{{\rm cut}}$, $E_{\bar\mu}\geq(E_{\bar\mu})_{{\rm cut}}$, and
$E_\gamma\geq(E_\gamma)_{{\rm cut}}$. Here, $E_\mu$, $E_{\bar\mu}$, and
$E_{\gamma}$ are the muon, anti-muon, and photon energies, respectively, in the
center of mass of the Higgs boson. For our present work, we restrict these cuts
to: $(m_{\mu\bar\mu}^2)_{{\rm cut}} \,, (m_{\mu\gamma}^2)_{{\rm cut}} \,,
(m_{\bar\mu\gamma}^2)_{{\rm cut}} \gg 4m_\mu^2$, and $(E_\mu)_{{\rm cut}} \,,
(E_{\bar\mu})_{{\rm cut}} \,, (E_\gamma)_{{\rm cut}}\gg 2m_\mu$.

 With these cuts imposed, the limits on the $dm_{\mu\gamma}^2$
and $dm_{\mu\bar\mu}^2$ integrations in Eqs.\,(\ref{dgam}) and (\ref{gam}) are
modified to
\begin{eqnarray}
(m_{\mu\gamma}^2)_{\rm min} & = &\;
{\rm max}[(m_{\mu\gamma}^2)_{{\rm cut}} \,, t_1]\,, \\
(m_{\mu\gamma}^2)_{\rm max} & = &\;
{\rm min}[t_2 \,, t_3]\,, \\
(m_{\mu\bar\mu}^2)_{\rm min} & = &\;
{\rm max}[(m_{\mu\bar\mu}^2)_{{\rm cut}} \,, s_1]\,,\\
(m_{\mu\bar\mu}^2)_{\rm max} & = &\;
{\rm min}[s_2  \,, s_3 ] \,,
\end{eqnarray}
where
\begin{eqnarray}
t_1 &=& 2m_H(E_\mu)_{{\rm cut}}- m_{\mu\bar\mu}^2\,, \\
t_2 &=& m_H^2-2m_H (E_{\bar\mu})_{{\rm cut}}\,, \\
t_3 &=& m_H^2-m_{\mu\bar\mu}^2-(m_{\bar\mu\gamma}^2)_{{\rm cut}}\,, \\
s_1 &=& 2m_H(E_\mu)_{{\rm cut}}+2m_H(E_{\bar\mu})_{{\rm cut}}-m_H^2   \,,\\
s_2 &=& m_H^2-2m_H (E_{\gamma})_{{\rm cut}}   \,,\\
s_3 &=& m_H^2-(m_{\mu\gamma}^2)_{{\rm cut}}-(m_{\bar\mu\gamma}^2)_{{\rm cut}}\,.
\end{eqnarray}
Notice that in order to implement and assign values to these cuts,
one needs to observe the following constraints
\begin{eqnarray}
(m_{\mu\bar\mu}^2)_{{\rm cut}}+ 2m_H (E_{\gamma})_{{\rm cut}} \leq m_H^2 \,,\\
(m_{\mu\gamma}^2)_{{\rm cut}}+ 2m_H (E_{\bar\mu})_{{\rm cut}} \leq m_H^2 \,,\\
(m_{\bar\mu\gamma}^2)_{{\rm cut}}+ 2m_H (E_{\mu})_{{\rm cut}} \leq m_H^2 \,,\\
(m_{\mu\bar\mu}^2)_{{\rm cut}}+(m_{\mu\gamma}^2)_{{\rm cut}}+
(m_{\bar\mu\gamma}^2)_{{\rm cut}} \leq m_H^2  \,,\\
(E_{\mu})_{{\rm cut}}+(E_{\bar\mu})_{{\rm cut}}+(E_{\gamma})_{{\rm cut}}\leq m_H \,,\\
(m_{\mu\bar\mu}^2)_{{\rm cut}} \,,\, (m_{\mu\gamma}^2)_{{\rm cut}} \,,\,
(m_{\bar\mu\gamma}^2)_{{\rm cut}} \gg 4m_\mu^2 \,,\\
(E_\mu)_{{\rm cut}} \,,\, (E_{\bar\mu})_{{\rm cut}} \,,\,
(E_\gamma)_{{\rm cut}} \gg 2m_\mu \,.
\end{eqnarray}

 In our present calculations for the decay width \linebreak
$\Gamma(H\rightarrow \mu\bar\mu\gamma)$ and the $\mu\bar\mu$-invariant mass
distribution $d\Gamma(H\rightarrow \mu\bar\mu\gamma)/dm_{\mu\bar\mu}$,
we choose the following set of cuts:
\begin{eqnarray}
(E_{\mu})_{{\rm cut}}=(E_{\bar\mu})_{{\rm cut}}=(E_{\gamma})_{{\rm cut}}= 1\, \rm{GeV} \,,\\
(m_{\mu\bar\mu}^2)_{{\rm cut}}=(m_{\mu\gamma}^2)_{{\rm cut}}=
(m_{\bar\mu\gamma}^2)_{{\rm cut}}= 25\, m_\mu^2 \,.
\end{eqnarray}

These cuts facilitate the experimental tagging of $\mu$, $\bar\mu$, and
$\gamma$. They provide minimum opening angles between $\mu$, $\bar\mu$, and
$\gamma$, and also avoid the collinear and infrared divergences. The cuts also
help discriminate the non-back-to-back $\mu\bar\mu$ pairs of the decay
$H\rightarrow \mu\bar\mu\gamma$ from the back-to-back $\mu\bar\mu$ pairs of the
decay $H\rightarrow \mu\bar\mu$. In principle, all the muons and photons of the
decays $H\rightarrow \mu\bar\mu\gamma$, $H\rightarrow \mu\bar\mu$, and
$H\rightarrow \gamma\gamma$ can be identified.

\section{Invariant Amplitudes}

\subsection{Tree Level Amplitudes}

 The leading order muon helicity non-flip amplitudes for the
decay $H\rightarrow \mu\bar\mu\gamma$ can be written as

\begin{equation} \label{treeamp_nf}
{\cal M}^{\rm tree}_{\lambda\bar{\lambda}\lambda_{\gamma}} =
i\frac{egm_{\mu}}{\sqrt{\displaystyle 2 t u}\,m_W}\left\{
\begin{array}{lcl}
+s &,& \lambda\bar{\lambda}\lambda_{\gamma} = --+ ,\\
-s &,& \lambda\bar{\lambda}\lambda_{\gamma} = ++- ,\\
+m_H^2 &,& \lambda\bar{\lambda}\lambda_{\gamma} = +++ ,\\
-m_H^2 &,& \lambda\bar{\lambda}\lambda_{\gamma} = --- ,\\
\end{array}
\right.\,
\end{equation}
where $\lambda=\pm1/2$, $\bar\lambda = \pm1/2$, and $\lambda_{\gamma}=\pm1$ are
the helicities of the $\mu$, $\bar\mu$, and $\gamma$, respectively and we show
only the signs of the helicities in Eq.\,(\ref{treeamp_nf}).

 The leading order muon helicity flip amplitudes are
\begin{eqnarray}\label{treeamp_f}
{\cal M}^{\rm tree}_{\lambda\bar{\lambda}\lambda_{\gamma}}
& = &i\frac{egm_{\mu}^2\sqrt{\displaystyle s}}{\sqrt{\displaystyle 2}\,m_W}
\frac{st+su+t^2+u^2}{(s+t)(s+u)} \nonumber\\ [4pt]
& &\times\left\{
\begin{array}{lcl}
1/t &,& \lambda\bar{\lambda}\lambda_{\gamma} = -++ ,\\
1/t &,& \lambda\bar{\lambda}\lambda_{\gamma} = +-- ,\\
1/u &,& \lambda\bar{\lambda}\lambda_{\gamma} = -+- ,\\
1/u &,& \lambda\bar{\lambda}\lambda_{\gamma} = +-+ .\\
\end{array}
\right.\
\end{eqnarray}

 In the Eqs.\,(\ref{treeamp_nf}) and
(\ref{treeamp_f}), we have kept only the leading orders in $m_{\mu}$, since we
are assuming $s, t, u \gg 4m_{\mu}^2$ and $E_{\mu}, E_{\bar\mu}, E_{\gamma}\gg
2m_{\mu}$. This is consistent with the cuts of $s, t, u \geq 25m_{\mu}^2 $ and
$E_{\mu}, E_{\bar\mu}, E_{\gamma}\geq 1$ GeV, that we impose on the
calculations of the decay width and the invariant mass distribution.

 It can be seen from the Eqs.\,(\ref{treeamp_nf}) and
(\ref{treeamp_f}) that the helicity flip amplitudes have an extra factor of
$m_{\mu}$. This is an expected behavior. As discussed in the
Ref.\,\cite{ab-cdr3}, the leptonic-current factors in the amplitudes, are
proportional to linear combinations of $\bar{u}(p_\mu)v(p_{\bar\mu})$ and
$\bar{u}(p_\mu)\sigma_{\alpha\beta}v(p_{\bar\mu})$. The muon helicity non-flip
contributions from these terms survive in the $m_{\mu}\rightarrow 0$ limit,
while the corresponding muon helicity flip contributions are proportional to
$m_\mu$, and vanish in this limit. For the tree level amplitudes, we may,
therefore, neglect the contribution from the muon helicity flip amplitudes, and
consider only the contribution from the muon helicity non-flip amplitudes to
the decay width and the invariant mass distribution.

\subsection{One-Loop Results}

 Contributions of the one-loop amplitudes to the decay
$H\rightarrow \mu\bar\mu\gamma$ arise from the diagrams illustrated in the
Fig.\,\ref{diag}. The explicit expressions for the amplitudes corresponding to
these diagrams are given in the Ref.\,\cite{ab-cdr1}. As discussed in the
Ref.\,\cite{ab-cdr3}, the leptonic-current factors in these amplitudes, are
proportional to linear combinations of
$\bar{u}(p_\mu)\gamma_{\alpha}v(p_{\bar\mu})$ and
$\bar{u}(p_\mu)\gamma_{\alpha}\gamma_5 v(p_{\bar\mu})$. The muon helicity flip
contributions from these terms survive in the $m_{\mu}\rightarrow 0$ limit,
while the muon helicity non-flip contributions, which are proportional to
$m_\mu$, do not. In this case, we may neglect the contribution from the muon
helicity non-flip amplitudes, and consider only the contribution from the muon
helicity flip amplitudes to the decay width and the invariant mass
distribution.

\begin{figure}
\hspace*{0.16in}
\epsfig{file=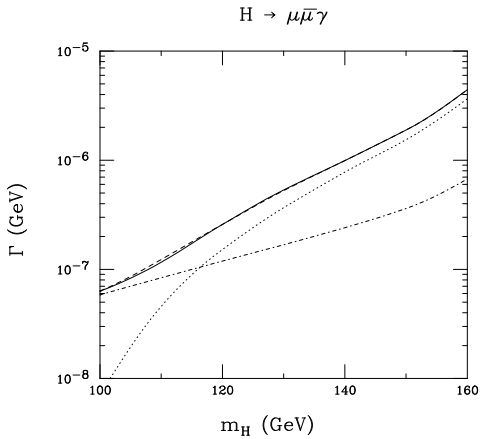,width=2.5in,angle=0}
\vspace{.10in}
\caption{The decay
width, at the loop level, for the decay mode $H\protect\rightarrow
\mu\bar\mu\gamma$ is shown. The solid line is the full loop calculation, the
dot-dashed line is the photon pole contribution, the dotted line is the $Z$
pole contribution, and the dashed line is sum of the photon and $Z$ pole
contributions.} \label{loop}
\end{figure}

 The expression for $\sum_{\rm spin}|{\cal M}|^2$, given
by Eq. (8) of the Ref.\,\cite{ab-cdr1}, together with Eq.\,(\ref{gam}), can be
used to calculate the one-loop contribution to $\Gamma(H\rightarrow
\mu\bar\mu\gamma)$. The results are illustrated in the Fig.\,\ref{loop}. In
this figure, contributions to the width from the triangle and box diagrams of
Fig.\,\ref{diag} are shown separately. The combined contributions from the $Z$
and photon poles in the Fig.\,\ref{diag}(a) constitute almost the entire
contribution of all diagrams in the Fig.\,\ref{diag}. Notice that for Higgs
boson masses not too much above 100 GeV, the photon pole makes substantial
contribution. Therefore, the simple estimate of the decay width
$\Gamma(H\rightarrow \mu\bar\mu\gamma)$, obtained by multiplying the width
$\Gamma(H\rightarrow Z\gamma)$ by the branching ratio $B(Z\rightarrow
\mu\bar\mu)$, will receive large correction due to the photon pole diagram.
However, for the $m_H \gtrsim 130$ GeV, it is the $Z$ pole that gives most of
the contribution.

\section{Tree and Loop Contributions}

 When combining the tree and one-loop contributions to obtain the muon helicity
flip and non-flip amplitudes, we can ignore the tree-one-loop interference
terms because of the suppression discussed in the previous section.
Consequently, the combined contributions from the tree and one-loop amplitudes
to the squared spin-summed amplitude, in the Eq.\,(\ref{dgam}),
\begin{equation}\label{sig1}
\sum_{\rm spin}|{\cal M}|^2 =
\sum_{\rm spin}|{{\cal M}^{\rm tree}_{\lambda\bar{\lambda}\lambda_{\gamma}}}
+ {{\cal M}^{\rm loop}_{\lambda\bar{\lambda}\lambda_{\gamma}}}|^2\,,\\
\end{equation}
can be simplified to
\begin{equation}\label{sig2}
\sum_{\rm spin}|{\cal M}|^2 =
\sum_{\rm spin}|{{\cal M}^{\rm tree}_{\lambda\bar{\lambda}\lambda_{\gamma}}}|^2 +
\sum_{\rm spin}|{{\cal M}^{\rm loop}_{\lambda\bar{\lambda}\lambda_{\gamma}}}|^2 \,,\\
\end{equation}
where
\begin{eqnarray}\label{sig3}
\sum_{\rm spin}|{{\cal M}^{\rm tree}_{\lambda\bar{\lambda}\lambda_{\gamma}}}|^2
& =& \;|{{\cal M}^{\rm tree}_{--+}}|^2+
|{{\cal M}^{\rm tree}_{++-}}|^2 \nonumber\\
&& +|{{\cal M}^{\rm tree}_{+++}}|^2+
|{{\cal M}^{\rm tree}_{---}}|^2
\,,\\[8pt]
\sum_{\rm spin}|{{\cal M}^{\rm loop}_{\lambda\bar{\lambda}\lambda_{\gamma}}}|^2
& =& \;|{{\cal M}^{\rm loop}_{-++}}|^2+
|{{\cal M}^{\rm loop}_{+--}}|^2 \nonumber\\
&& +|{{\cal M}^{\rm loop}_{+-+}}|^2+
|{{\cal M}^{\rm loop}_{-+-}}|^2 \,.
\end{eqnarray}

 Using Eq.\,(\ref{sig2}), we may write the following relations for
the $\mu\bar\mu$-invariant mass distributions and the decay widths
\begin{equation}\label{dgamtotal}
\frac{d\Gamma}{dm_{\mu\bar\mu}} = \frac{d\Gamma^{\rm{tree}}}{dm_{\mu\bar\mu}}
+ \frac{d\Gamma^{\rm{loop}}}{dm_{\mu\bar\mu}} \,,\\
\end{equation}
\begin{equation}\label{gamtotal}
\Gamma = \Gamma^{\rm{tree}} + \Gamma^{\rm{loop}} \,.\\
\end{equation}

\begin{figure}
\hspace*{0.16in}
\epsfig{file=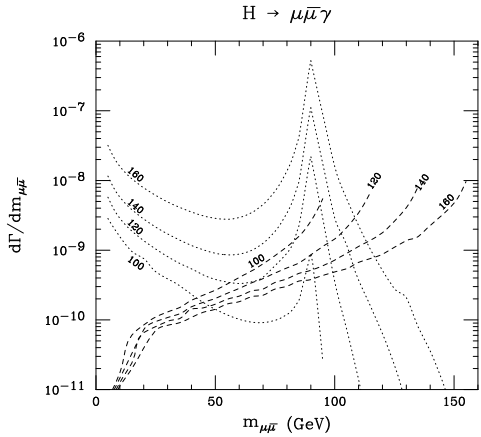,width=2.5in,angle=0}
\vspace{.10in}
\caption{The
$\mu\bar\mu$-invariant mass distribution of the decay $H\protect\rightarrow
\mu\bar\mu\gamma$, for Higgs boson masses of 100 GeV, 120 GeV, 140 GeV, and 160
GeV is shown. The dotted lines are for the loop contribution and dashed lines
are for the tree level contribution. The combined contribution is simply the
sum of the tree and the loop contributions.} \label{distn}
\end{figure}

\begin{figure}
\hspace*{0.16in}
\epsfig{file=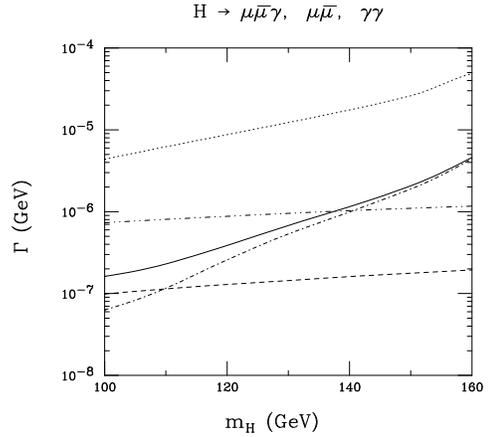,width=2.5in,angle=0}
\vspace{.10in}
\caption{The decay
width for several decay modes of the Higgs boson is shown. The dashed line is
the width for the decay $H\protect\rightarrow \mu\bar\mu\gamma$ at tree level,
the dot-dashed line is the width at the one-loop level, the solid line is the
total width (the sum of the dashed and dot-dashed lines), the dot-dot-dashed
line is $\Gamma(H\protect\rightarrow \mu\bar\mu)$ at tree level, and the dotted
line is $\Gamma(H\protect\rightarrow \gamma\gamma)$.} \label{width}
\end{figure}

 In Fig.\,\ref{distn}, we show $\mu\bar\mu$-invariant mass distributions for the decay
$H\rightarrow \mu\bar\mu\gamma$. This figure illustrates the tree and one-loop
contributions to the invariant mass distributions
$d\Gamma^{\rm{tree}}/dm_{\mu\bar\mu}$ and
$d\Gamma^{\rm{loop}}/dm_{\mu\bar\mu}$, respectively. The complete distribution,
$d\Gamma/dm_{\mu\bar\mu}$, according to the Eq.\,(\ref{dgamtotal}), is simply
the sum of these two contributions. Fig.\,\ref{width}, shows the tree and
one-loop contributions, $\Gamma^{\rm{tree}}$ and $\Gamma^{\rm{loop}}$, as well
as their sum, $\Gamma$. For comparison, the widths of the tree level decay
$H\rightarrow \mu\bar\mu$ the decay $H\rightarrow \gamma\gamma$ are included.

\section{Summary}

 We have shown that, for $m_H \lesssim 130$ GeV, the contributions
from the tree and loop levels to the decay $H\rightarrow \mu\bar\mu\gamma$ are
comparable. It is, therefore, necessary to include both contributions to the
$\mu\bar\mu$-invariant mass distribution and the decay width. In this mass
region, as elsewhere, the interference terms between the tree and loop
amplitudes are small compared to the squared amplitude of either one, and the
total decay width is simply the sum of the decay widths from the tree and
one-loop contributions, $\Gamma = \Gamma^{\rm{tree}} + \Gamma^{\rm{loop}}$.
When $m_H \gtrsim 130$ GeV, the one-loop contribution dominates, and, for $m_H$
larger than 140 GeV, it exceeds the tree level contribution to
$H\to\mu\bar{\mu}$.

\begin{figure}
\hspace*{0.16in}
\epsfig{file=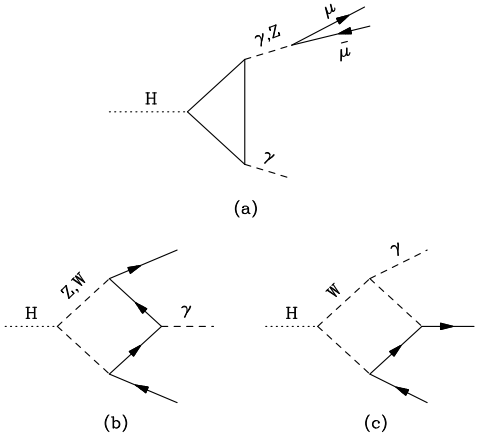,width=2.5in,angle=0}
\vspace{.10in}
\caption{The diagrams
for the decay mode $H\rightarrow \mu\bar\mu\gamma$ at the loop level are
shown.} \label{diag}
\end{figure}

\begin{figure}
\hspace*{0.16in}
\epsfig{file=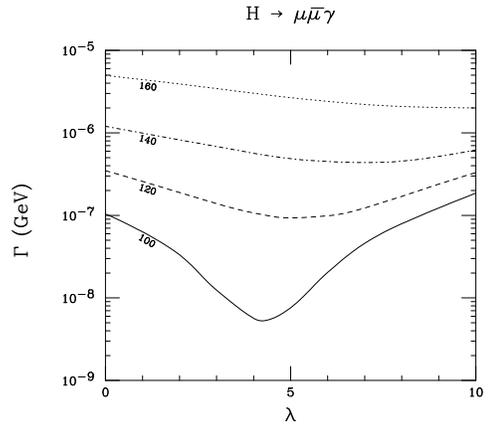,width=2.5in,angle=0}
\vspace{.10in}
\caption{The decay
width $\Gamma(H\protect\rightarrow \mu\bar\mu\gamma)$, at the loop level, as a
function of the $Ht\bar{t}$ coupling (in multiples $\lambda$ of the Standard
Model coupling), for several values of Higgs boson mass is shown. The solid
line is $m_H=100$ GeV, the dashed line is $m_H=120$ GeV, the dot-dashed line is
$m_H=140$ GeV, and the dotted line is $m_H=160$ GeV.} \label{ttbar}
\end{figure}

 Finally, the presence of the top quark loop in some of the diagrams in
the Fig.\,\ref{diag}(a) offers an opportunity to use the decay $H\rightarrow
\mu\bar\mu\gamma$ as a possible probe of the Higgs boson coupling to the top
quark. In Fig.\,\ref{ttbar}, we show the decay width for $H\rightarrow
\mu\bar\mu\gamma$ that arises from the complete set of diagrams in
Fig.\,\ref{diag} after modifying the $Ht{\bar{t}}$ coupling. Modification has
been achieved by multiplying the Standard Model $Ht{\bar{t}}$ coupling by a
factor $\lambda$ \cite{hks}.

 As it can be seen from the Fig.\,\ref{ttbar}, the higher
the Higgs boson mass is, the more difficult it is to distinguish the Standard
Model coupling for $Ht{\bar{t}}$, $\lambda = 1$, from the case of no coupling
between the Higgs boson and the top quark, $\lambda = 0$. Generally, a measured
value of the decay width $\Gamma$ corresponds to two different values of
$\lambda$. Therefore, the determination of $\lambda$ will not be unique.
However, it can be used as an indication of a deviation from the Standard
Model.

\acknowledgments
One of us (A.A.) wishes to thank the Department of Physics and Astronomy
at Michigan State University for its hospitality and computer resources.
This work was supported in part by the National Science Foundation
under Grant No. PHY-9802439.

\end{document}